\documentclass[%
 reprint,
superscriptaddress,
 amsmath,amssymb,
 aps,
floatfix,
]{revtex4-2}

\usepackage{graphicx}
\usepackage{dcolumn}
\usepackage{bm}
\usepackage{scalerel}
\usepackage{textcomp}
\usepackage{amsmath}
\usepackage{physics}
\usepackage[caption=false]{subfig}
\usepackage{booktabs}

\usepackage{appendix}

\newcommand{\textsub}[1]{$_{\text{#1}}$}

\newcommand{\figref}[2]{\hyperref[#1]{\ref{#1}(#2)}}
\newcommand{\figsref}[2]{\hyperref[#1]{\ref{#1}(#2)}}

\usepackage{xcolor}
\definecolor{darkblue}{RGB}{0,0,150}
\definecolor{nightblue}{RGB}{0,0,100}
\usepackage[
bookmarksnumbered,
pdfpagelabels=false,
colorlinks,
linkcolor=nightblue,
citecolor=nightblue,
filecolor=black,
urlcolor=darkblue,
breaklinks=true
]{hyperref}
\begin{document}

\preprint{APS/123-QED}

\title{Chiral Majorana Modes via Proximity \\ to a Twisted Cuprate Bilayer}

\author{Gilad Margalit}
 \affiliation{Department of Condensed Matter, Weizmann Institute of Science, Rehovot, Israel 7610001}
\author{Binghai Yan}
 \affiliation{Department of Condensed Matter, Weizmann Institute of Science, Rehovot, Israel 7610001}
 \author{Marcel Franz}
 \affiliation{Department of Physics and Astronomy, and Quantum Matter Institute, University of British Columbia, Vancouver, British Columbia, Canada V6T 1Z1}
\author{Yuval Oreg}
 \affiliation{Department of Condensed Matter, Weizmann Institute of Science, Rehovot, Israel 7610001}

\date{\today}

\begin{abstract}
We propose a novel heterostructure to achieve chiral topological superconductivity in 2D. A substrate with a large Rashba spin-orbit coupling energy is brought in proximity to a twisted bilayer of thin films exfoliated from a high-temperature cuprate superconductor. The combined system is then exposed to an out-of-plane magnetic field. The rare $d + id$ pairing symmetry expected to occur in such a system allows for nontrivial topology; specifically, in contrast to the case of the twisted bilayer in isolation, the substrate induces an odd Chern number. The resulting phase is characterized by the presence of a Majorana zero mode in each vortex.
\end{abstract}

\maketitle

\section{Introduction}
\label{Introduction}

Following the discovery by Yu \textit{et al.} that the cuprate superconductor (SC) Bi$_2$Sr$_2$CaCu$_2$O$_{8+\delta}$ (Bi-2212) can be exfoliated to produce a monolayer that retains a large critical temperature $T_C$ \cite{Yu2019}, the possibilities have expanded for topological superconductivity in 2D. Notably, it was recently proposed by Can \textit{et al.} \cite{Can2021} to combine this development with the recent paradigm of twisted bilayer graphene \cite{Cao2018,Cao2018a} and realize a topological superconductor (TSC) by stacking cuprate monolayers with a twist angle close to 45$^\circ$. As we will review in Sec.~\ref{Bilayer}, the layers' $d$-wave pairing symmetry and spontaneous breaking of time reversal symmetry yield an effective $d + id$ superconductor, which for sufficiently strong interlayer coupling can exhibit nontrivial topology. This setup was tested shortly after by Zhao \textit{et al.} \cite{Zhao2021}, in which the co-tunneling behavior of Cooper pairs suggested a TSC phase. 

\begin{figure}
     \includegraphics[width=0.4\textwidth]{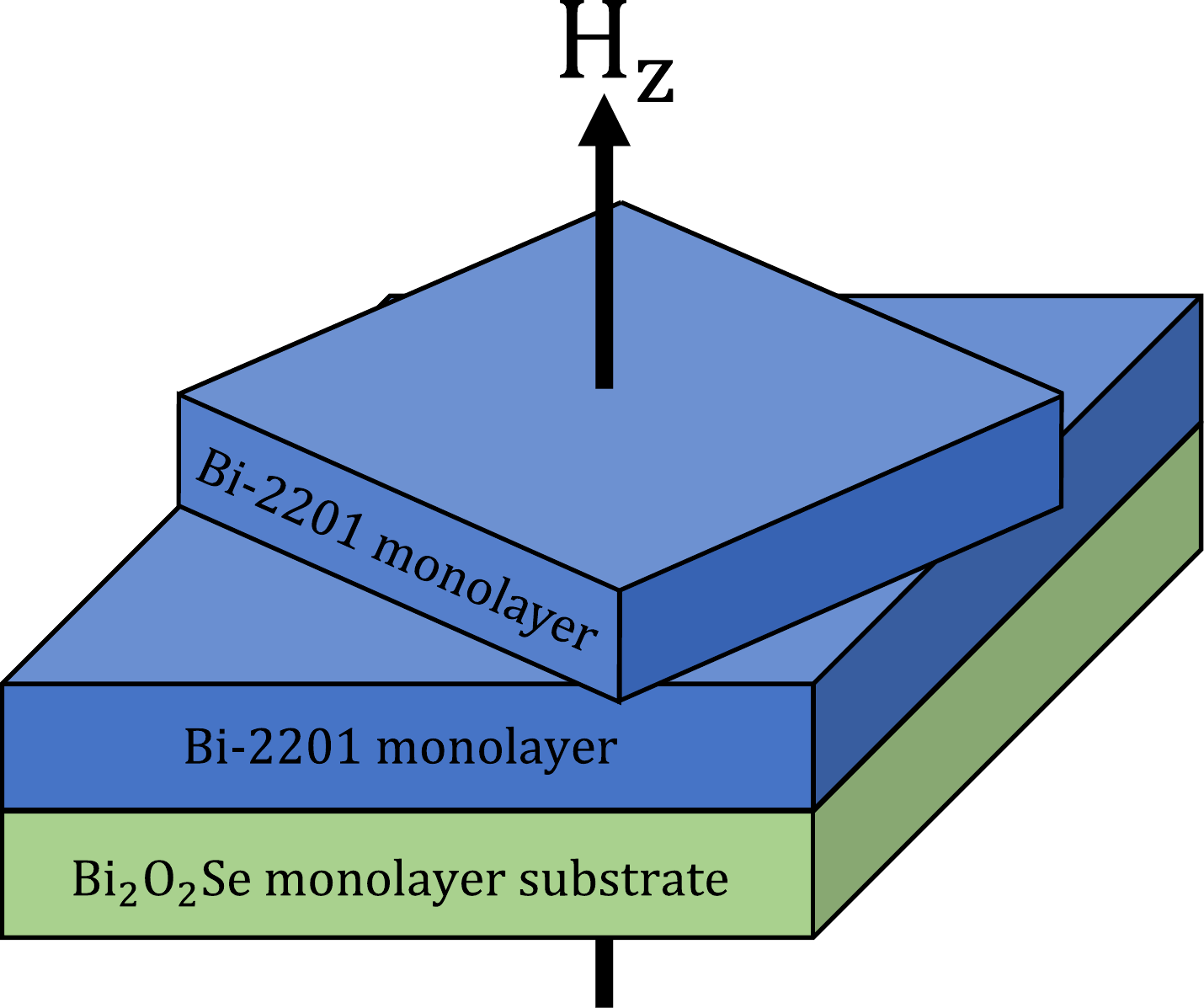}
        \caption{Schematic representation of our proposed heterostructure. Two monolayer cuprates (in this case layers of Bi-2201) are twisted close to 45$^\circ$ with respect to each other, and the bottom layer is aligned with a substrate with similar lattice constant (in this case a monolayer of Bi$_2$O$_2$Se). The substrate contributes a strong Rashba spin-orbit coupling, while the bilayer contributes $d + id$-wave pairing. Then, the structure is placed in an out-of-plane magnetic field $H_z$.}
    \label{layersSchematic}
\end{figure}

However, the Chern numbers obtained by Can \textit{et al.} are all even, indicating that the chiral edge modes hosted on each vortex always come in pairs and are thus not required to include a zero-energy state. For this reason, we propose a variation on this heterostructure design, shown in Fig.~\ref{layersSchematic}. Our heterostructure consists of a cuprate bilayer with a near-45$^\circ$ twist, similar to that discussed by Can \textit{et al}, but with the addition of a proximity-coupled conducting substrate. The substrate must have a large Rashba spin-orbit coupling (SOC) energy relative to its bandwidth. In this work, we propose that this modified system, when subjected to an out-of-plane magnetic field, can yield a topological phase of Chern number $\pm 1$. This phase includes a single chiral Majorana mode in each vortex produced by the magnetic flux through the system. 

In Sec.~\ref{Model}, we outline a simple Hamiltonian which describes our system, and show that, with proper gating of the chemical potential, the chiral phase can be achieved for arbitrarily small values of the magnetic field $H_z$. This is crucial, as the energy associated with the field, which is of the order $\mu_B H_z$ for the Bohr magneton $\mu_B$, is much smaller than the energy scales relevant to the band structure (and in fact, since we are using a high-temperature superconductor, the relevant superconducting gap can also exceed this value). This property is present for a $d + id$-wave superconductor but not a standard $s$-wave, and is the reason that proximity coupling to an $s$-wave will not suffice.

We conclude our analytical justification in Sec.~\ref{Bilayer}, where we briefly summarize the conclusion of Can \textit{et al.} that a twisted bilayer of monolayer cuprate superconductors will form an effective $d + id$-wave pairing symmetry. The superconducting pairing term inherited in the substrate from the proximity effect must therefore also have $d + id$-wave character (we use a Hubbard-Stratonovich transformation to prove this in Appendix~\ref{derivation}).

To illustrate the plausibility of our proposed heterostructure, we then describe a first-principles calculation we conducted with density functional theory (DFT) in Sec.~\ref{DFT}. We simulate a bilayer of the high-$T_C$ superconductor Bi$_2$Sr$_2$CuO$_{6+\delta}$ (Bi-2201), including the effects of ionic relaxation, and demonstrate coupling between the layers as well as with the chosen substrate, a monolayer of the semiconductor Bi$_2$O$_2$Se \cite{Wu2017}. This substrate, in addition to having similar composition and a matching lattice constant to the bilayer (both of which tend to enhance coupling), is shown to have a conducting band with strong SOC which can be reached by gating the system. We show that states associated with the substrate extend into the bilayer, demonstrating the large coupling between bilayer and substrate.

Parameters obtained from our simulation allow us to estimate the resulting superconducting gap in Sec.~\ref{SC} to be of order 0.01 or 0.1 meV. At temperatures below this gap, we expect to see chiral Majorana modes bound to each vortex in the heterostructure, which exhibit non-Abelian exchange statistics \cite{Nayak2008}. Since our computed pairing gaps are comparable to the temperatures of other TSC experiments~\cite{Mourik2012}, this system shows promise as a realizable TSC.

\section{Model}
\label{Model}

Following Sato \textit{et al.} \cite{Sato2010a}, we describe this system's Hamiltonian $\mathcal{H}$ with a phenomenological tight-binding model expressed in the Bogoliubov-de-Gennes (BdG) formalism. The non-superconducting parameters of this model will then be determined via our DFT simulation. 
\begin{equation}
\label{Hamiltonian_bilayer}
\begin{aligned}
\mathcal{H} &= \mathcal{H}_\text{0} + \mathcal{H}_\text{SC} + \mathcal{H}_\text{Z} + \mathcal{H}_\text{SOC}; \\
\mathcal{H}_\text{0} &= \sum_{\bm{k},s}{\epsilon_{\bm{k}} c_{\bm{k}s}^\dagger c_{\bm{k}s}} \\
\mathcal{H}_\text{SC} &= \sum_{\bm{k}}{( \Delta_{\bm{k}}c_{\bm{k}\uparrow}^\dagger c_{-\bm{k}\downarrow}^\dagger - \Delta_{\bm{k}}^*c_{\bm{k}\uparrow}c_{-\bm{k}\downarrow})} \\
\mathcal{H}_\text{Z} &= \mu_B H_z \sum_{\bm{k},s}{c_{\bm{k}s}^\dagger \sigma_{s s'}^z c_{\bm{k}s'}} \\
\mathcal{H}_\text{SOC} &= \alpha \sum_{\bm{k},\sigma}{c_{\bm{k}s}^\dagger(\sin k_xa \: \sigma_{s s'}^y - \sin k_ya \: \sigma_{s s'}^x) c_{\bm{k}s'}},
\end{aligned}
\end{equation}
where $\mathcal{H}_\text{0}$ is a spin-degenerate pair of ordinary substrate bands, which we assume to have the simple form
\begin{equation}
\label{ek}
    \epsilon_{\bm{k}} = -\mu-2t(\cos{k_ya}+\cos{k_xa})
\end{equation}
for chemical potential $\mu$ and bandwidth $4t$. $\mathcal{H}_\text{SC}$ contains the $d + id$-wave order parameter
\begin{equation}
\label{deltak}
    \Delta_{\bm{k}} = \frac{\Delta}{2} \Big[(\cos{k_ya_0}-\cos{k_xa_0}) + 2i\sin{k_xa_0}\sin{k_ya_0}\Big]
\end{equation}
proportional to $\Delta$, the maximum pairing gap. Finally, we add a Zeeman term $\mathcal{H}_\text{Z}$ and Rashba SOC $\mathcal{H}_\text{SOC}$.

In the above equations, $\sigma^i$ are Pauli matrices, $a$ is the moir\'e cell lattice constant, and $a_0$ is the superconductor's unit cell lattice constant ($a = \sqrt{5}a_0$ for our implementation). We will use a subscript 0 to refer to parameters at the scale of the unit cell (the SC bilayer and substrate have, to good approximation, the same lattice constant). Though the moir\'e supercell introduces several folded pairing terms, it is sufficient to only consider the terms in Eq.~\eqref{deltak} since only a single, small-momentum Fermi pocket will contribute, and all other pairing terms are not close to zero-momentum at the Fermi energy.

\begin{figure}
     \subfloat[]{\includegraphics[width=0.25\textwidth]{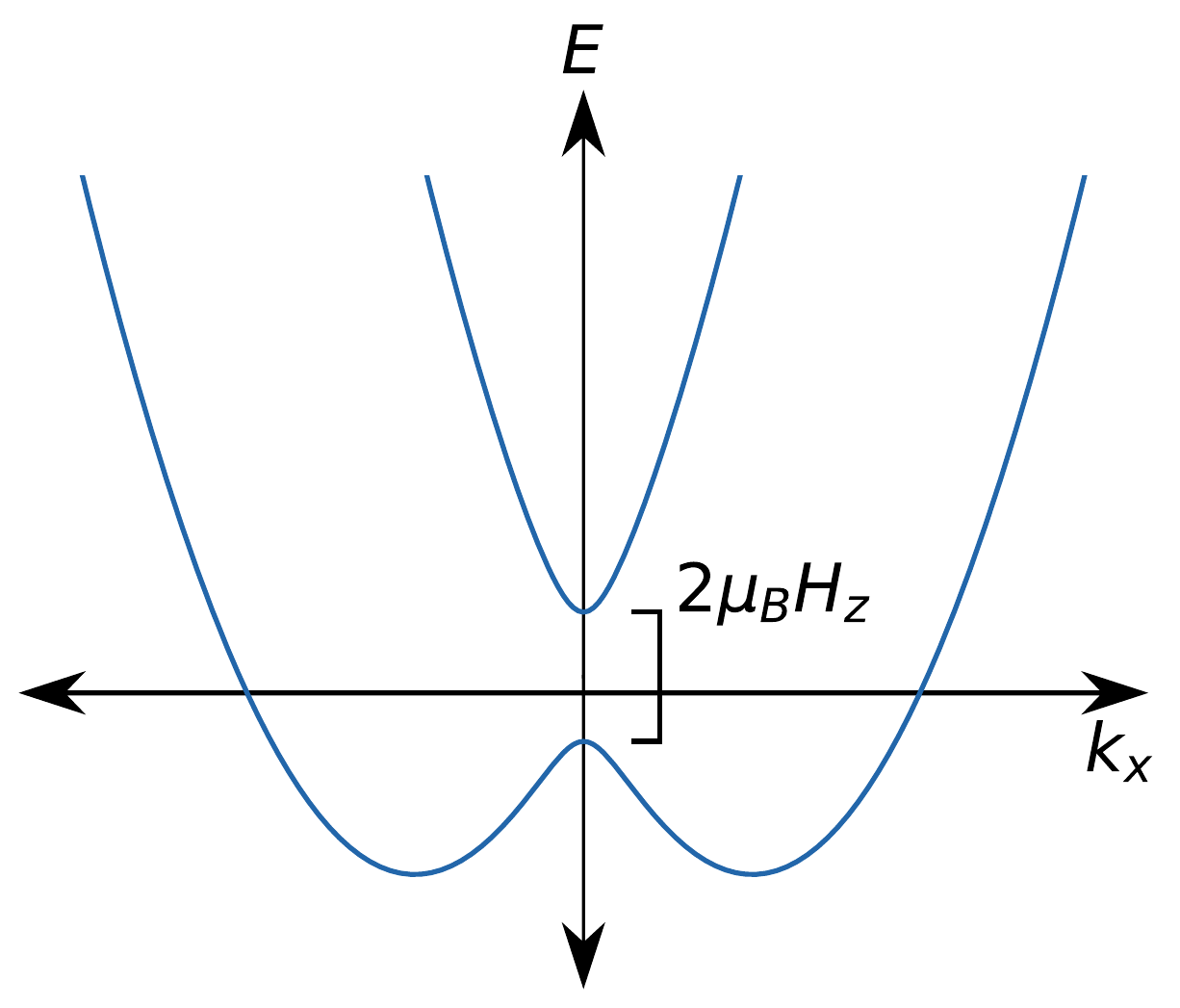}
         \label{schematicHz}} \\
     \subfloat[$s$-wave]{\includegraphics[width=0.24\textwidth]{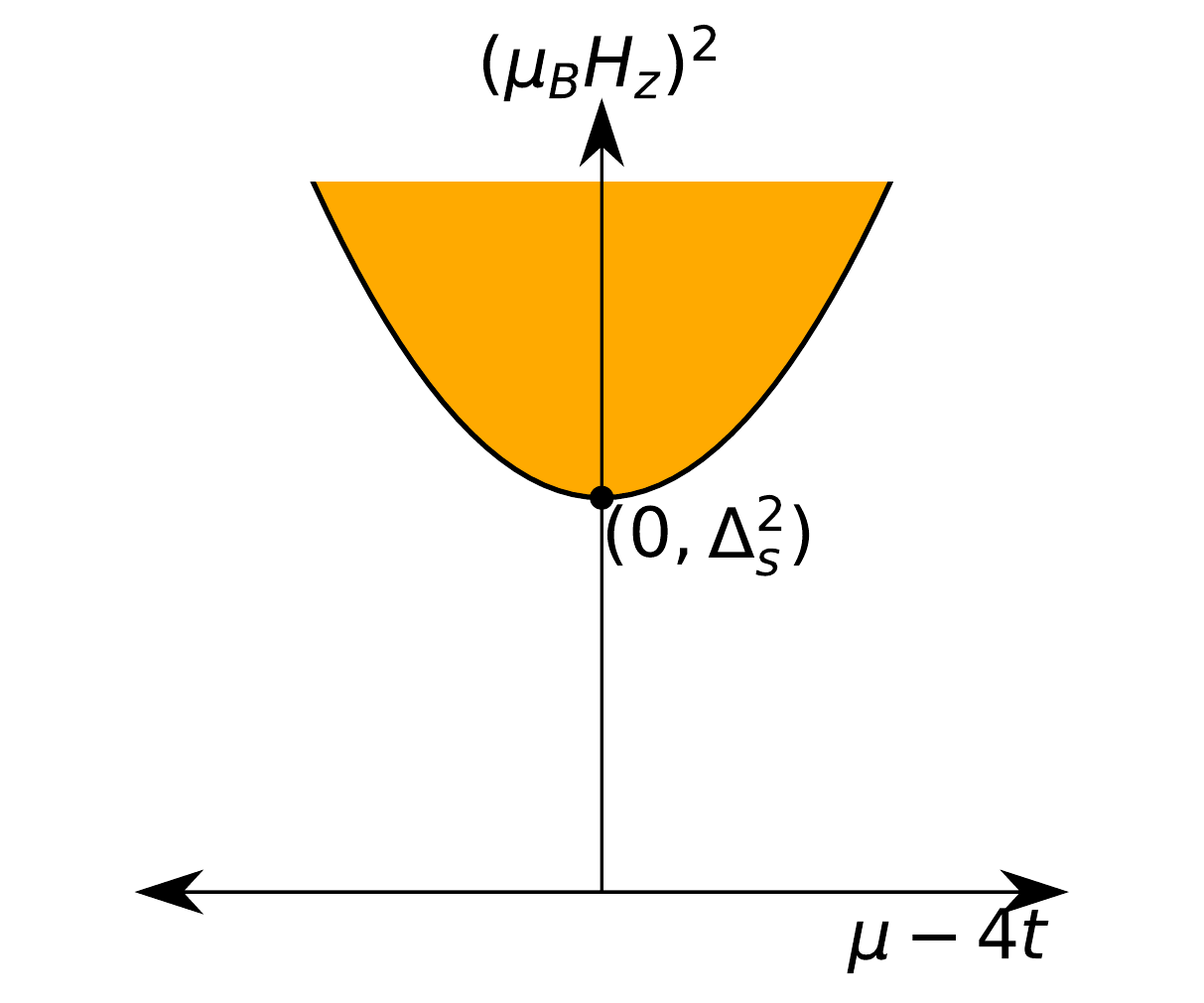}
         \label{SatoSwave}}
     \subfloat[$d + id$-wave]{\includegraphics[width=0.24\textwidth]{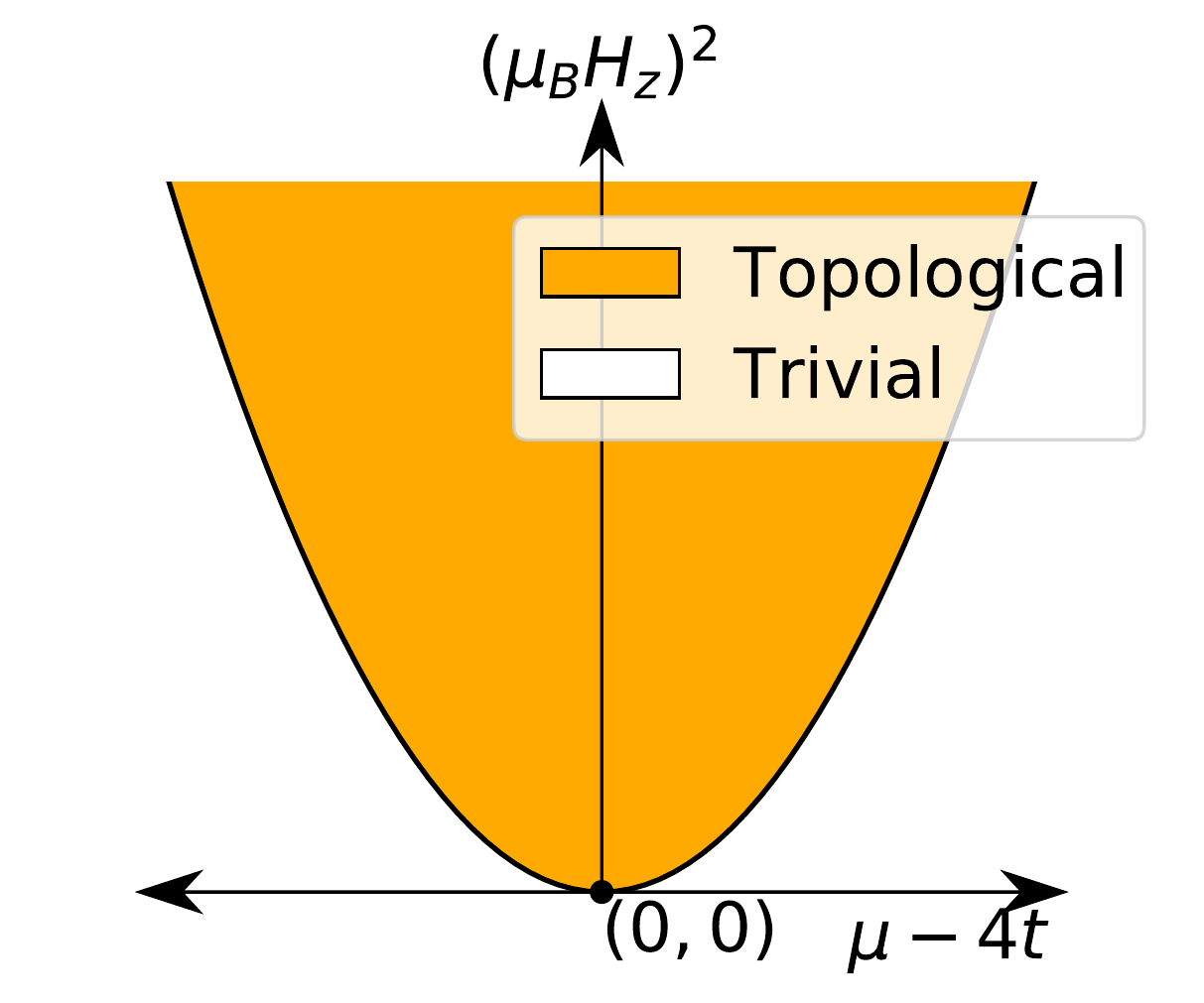}
         \label{SatoDwave}} \\
         
        \caption{(a) Schematic band structure without superconductivity for $k_y = 0$. A topological phase requires $\sqrt{(\mu-4t)^2 + \Delta_{\bm{k}=0}^2} < \abs{\mu_B H_z}$, where $\mu - 4t$ is the bands' energy at zero momentum in the absence of the external field $H_z$. In effect, the Fermi energy ($E = 0$) must lie within the gap opened by $H_z$. (b) and (c) Topological phase diagram for s-wave and $d + id$-wave superconductors with Rashba SOC and an applied magnetic field, respectively. In the s-wave case, there can only be an odd Chern number (gold region) if $\abs{\mu_B H_z} > \abs{\Delta_s}$ (where $\Delta_s$ is the s-wave pairing energy), while in the $d + id$-wave case, there is no lower bound for $H_z$. This is because the $d + id$-wave pair potential is gapless at $\bm{k}=0$, and hence there is no pairing energy for the external field to overcome.}
        \label{PhaseDiagrams}
\end{figure}

Fig.~\figref{PhaseDiagrams}{a} shows a schematic band structure of just the electron part of the Hamiltonian (i.e. without $\mathcal{H}_\text{SC}$) for $k_y = 0$. As long as $\mu$ is tuned via gating such that the Fermi energy lies within the gap opened by the Zeeman field $H_z$, the system is in its topological phase.

As shown in Figs.~\figsref{PhaseDiagrams}{b} and \figsref{PhaseDiagrams}{c}, this would not necessarily be the case if our superconductor were $s$-wave; in such systems, the Zeeman energy must exceed the pairing energy in order to result in the TSC phase \cite{Sato2010a}. However, the $d+id$ character of the bilayer SC works to our advantage. Since the $d + id$ pair potential is gapless at the $\Gamma$ point, while the $s$-wave potential has a constant gap of $\Delta_s$, the $d + id$-wave can enter the TSC phase for arbitrarily small values of $H_z$. This is especially important in this case, since the maximum pairing energy for a Bi-2201 monolayer may be of order 10 meV \cite{Kugler2001}, which is too high a threshold to be practical for an external field.

Sato \textit{et al.} proved that for this system, the TSC phase has Chern number 1. To give intuition for their result, this case is closely analogous to the case of an $s$-wave with Rashba SOC in a Zeeman field, where the field opens a gap by coupling the in-plane spin components of a single Fermi pocket \cite{Oreg2010}. This results in an effective spinless $p$-wave, an archetypical chiral TSC with odd Chern number. Since this argument depends only on finite pairing along the single Fermi pocket, it works equally well for $d + id$-wave pairing as for $s$-wave.

It is important to note that while the cuprate bilayer has an intrinsic spin-orbit interaction that is not of the Rashba form (and that is not included in the model above), this is not sufficient (when combined with an out-of-plane Zeeman field) to open the gap shown in Fig.~\figref{PhaseDiagrams}{a}. Rather, a substrate, and its associated Rashba SOC, is essential. This is because in the absence of the substrate, the symmetries $M_xM_zT$ and $M_yM_zT$ are conserved, where $M_i$ is a mirror symmetry along the $i$-axis and $T$ is time-reversal symmetry. (Here, $x$ and $y$ are the horizontal and vertical directions of Fig.~\ref{CuLayers}, respectively, not aligned with either lattice, while $z$ is normal to the plane.) Though the Zeeman field and the $d + id$ pairing both break time-reversal symmetry alone, both preserve these combined symmetries. Any intrinsic SOC of the bilayer must conserve these symmetries as well, since all electronic structure terms have the same symmetry constraints as their material's atomic structure. This enforces a band crossing at the $\Gamma$ point, and thus the gap in Fig.~\figref{PhaseDiagrams}{a} remains closed.

These conserved symmetries are the underlying reason why odd Chern numbers are not possible in the bilayer alone. When a substrate is placed on the $-z$ side of the bilayer but not on the $+z$ side, however, it breaks $M_z$, and thus also breaks $M_xM_zT$ and $M_yM_zT$, allowing for an odd-Chern TSC gap to form.

\section{Twisted Bilayer with $\mathbf{d+id}$ Superconductivity}
\label{Bilayer}

In this section, we briefly summarize the argument by Can \textit{et al.} \cite{Can2021} that a cuprate bilayer with a twist angle close to 45$^\circ$ results in a $d + id$-wave system in the regime of strong coupling between the layers. This spontaneous time-reversal breaking was further supported by an incoherent tunnelling model in a follow-up work \cite{Haenel2022} and used as the basis of a similar proposal to this work, which uses a topological insulator substrate \cite{Mercado2022}.

Bulk Bi-2201 is a layered high-temperature superconductor with a large pairing gap of approximately 12 meV at its widest point \cite{Fischer2007}. If, like Bi-2212, it retains its strong superconductivity in the monolayer limit, it would be a very promising candidate for a TSC heterostructure. To good approximation, it has a $d_{x^2-y^2}$ order parameter (which can be seen in angle-resolved ARPES measurements \cite{Damascelli2003}), so its superconducting gap closes along two orthogonal nodal lines in the Fermi surface.

\begin{figure}
     \includegraphics[width=0.3\textwidth]{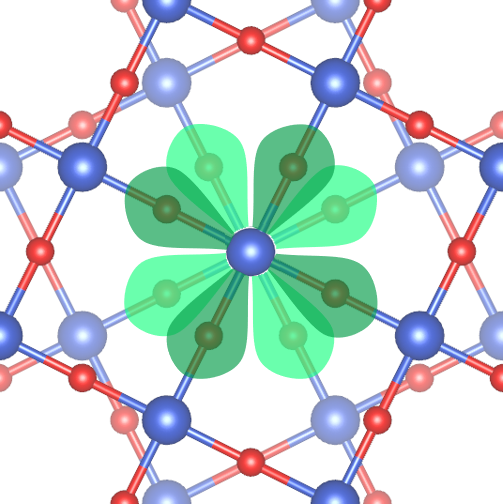}
    \caption{Bilayer lattice showing only the Cu-O layers for clarity (before ionic relaxation is taken into account). The superconducting order parameter is represented schematically in green for each layer. Since the layers are twisted with a relative angle close to 45$^\circ$ (36.9$^\circ$ in this case, to yield a finite-size moir\'e cell), the $d_{x^2-y^2}$ order parameter in the bottom (lighter) layer resembles a $d_{xy}$ order parameter in the coordinate frame of the top (darker) layer. When time reversal symmetry is spontaneously broken, this results in effective overall $d + id$-wave superconductivity in the bilayer.}
    \label{CuLayers}
\end{figure}

However, as illustrated in Fig.~\ref{CuLayers}, when two layers are twisted with a relative angle close to 45$^\circ$ (36.9$^\circ$ in this case, to yield a finite-size moir\'e cell), the $d_{x^2-y^2}$ order parameter of one layer acts as a $d_{xy}$ order parameter in the coordinate frame of the other, allowing for one order parameter to be finite along the nodal lines of the other.

The combined pair potential has the form
\begin{equation}
\begin{aligned}
\label{Deltakcombined}
    \Delta_{\bm{k}} \sim &\Big[\cos 2\theta (\cos{k_ya_0}-\cos{k_xa_0}) \\
    &+ 2\sin 2\theta e^{i\phi}\sin{k_xa_0}\sin{k_ya_0}\Big]
\end{aligned}
\end{equation}
for twist angle $\theta$ and a relative phase $\phi$ between the order parameters of the two layers. The phase $\phi$ can be determined from the Ginzburg-Landau description of the free energy of the bilayer. Up to second order in the order parameters for the two layers, $\psi_1 \equiv \psi$ and $\psi_2 = \psi e^{i\phi}$, C\textsub{4} symmetry constrains the free energy to have the form
\begin{equation}
\label{GL}
    \mathcal{F}(\phi) = \mathcal{F}_0 + 2\psi^2\big[-B_0\cos 2\theta \cos\phi + C\psi^2\cos 2\phi\big],
\end{equation}
where $B_0$ and $C$ are real constants representing single-pair and two-pair tunnelling between the layers, respectively. Up to an overall scale factor, $C$ should therefore be proportional to the square of $B_0$, and thus we expect $C$ to be positive. It is easy to show that under these constraints, at $\theta = 45^\circ$, the free energy has minima at $\phi = \frac{\pi}{2}$ and $\phi = -\frac{\pi}{2}$, and hence a $d + id$ or $d - id$ order parameter is energetically favorable. Angles offset from 45$^\circ$ also result in a phase of $\pm \frac{\pi}{2}$ as long as the coupling between the layers is sufficiently large to overcome the offset. Without loss of generality, we will assume a $d + id$-wave order parameter rather than a $d - id$-wave (the only relevant difference between the two cases is that the Chern number changes sign). As Can \textit{et al.} \cite{Can2021} showed, this order parameter can give rise to topological phases of Chern number 0, 2, or 4. However, without our proposed addition of a substrate and external field, it cannot result in odd-Chern phases, which host unpaired chiral Majorana zero modes.

The pairing process occurs within the cuprate bilayer; however, as we prove in Sec. \ref{SC}, the $d + id$ nature of the pairing is also adopted by the substrate via the proximity effect. All that remains is to show that there can exist a Rashba pair of substrate bands which can couple strongly to the Bi-2201 bilayer (so as to achieve a large effective pairing from the proximity effect). In the next section, we use DFT to suggest that such a band pair is indeed present.

\section{First-Principles Calculation}
\label{DFT}

We used DFT to simulate a moir\'e cell of the twisted Bi-2201 bilayer and monolayer Bi$_2$O$_2$Se substrate. This particular combination was chosen because Bi-2201 is the simplest high-$T_C$ layered cuprate, and Bi$_2$O$_2$Se had excellent lattice constant matching and similar composition to Bi-2201, which aids in coupling between the substrate and bilayer. We employed the Vienna \textit{Ab-initio} Simulation Package (VASP)\cite{Kresse1993, Kresse1999} with generalized gradient approximation (GGA) functionals \cite{GGA} and included an ionic relaxation step after the materials were brought in contact.

\begin{figure}
     \subfloat[]{\includegraphics[width=0.24\textwidth]{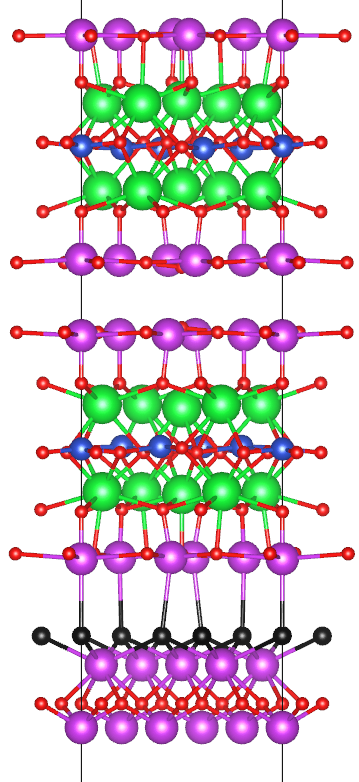}
         \label{unitCell}}
     \subfloat[]{\includegraphics[width=0.224\textwidth]{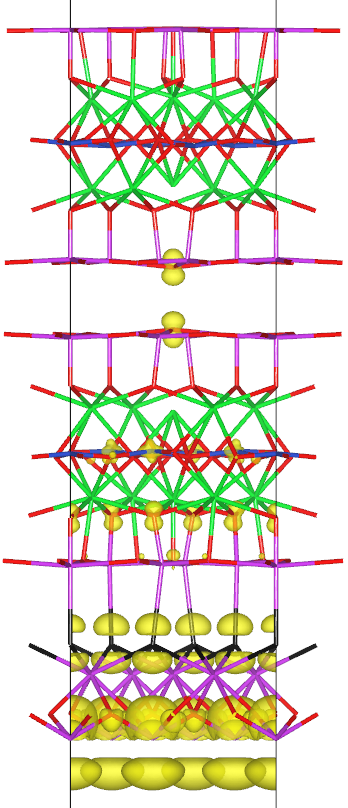}
         \label{chgDensity}}
         
        \caption{(a) Crystal structure of our proposed heterostructure. A full moir\'e cell (5 unit cells) is shown, with two twisted Bi-2201 layers above a Bi$_2$O$_2$Se monolayer. (b) Isosurface of the combined charge density (yellow) at $\Gamma$ of the two highlighted substrate bands in Fig.~\ref{bilayerBands}. Atoms have been omitted for clarity. Note that while electron density is mainly located in the substrate, it extends into the middle of the bilayer, suggesting good coupling and hence a strong proximity effect.}
        \label{structure}
\end{figure}

The moir\'e cell had a twist angle of 36.9$^{\circ}$, leading to exactly 5 unit cells per moir\'e cell (the same angle used by Can \textit{et al.} \cite{Can2021}); this is close enough to 45$^{\circ}$ to result in a $d + id$-wave superconductor while having the smallest-possible moir\'e cell for ease of computation. Fig. \figref{structure}{a} shows the moir\'e cell, with Bi in purple, Ca in green, Cu in blue, O in red, and Se in black.

\begin{figure}
     \includegraphics[width=0.45\textwidth]{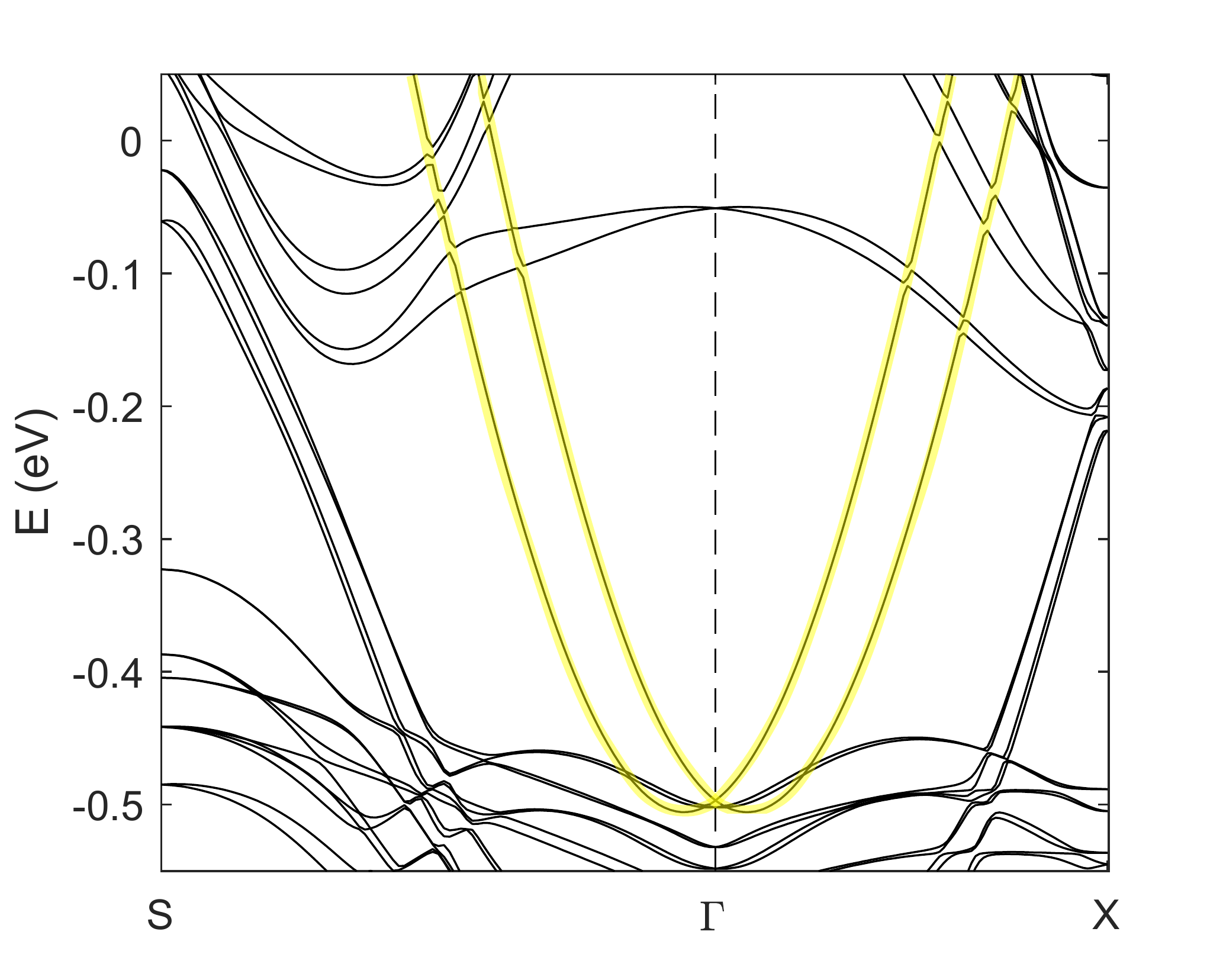}
        \caption{Computed band structure of the twisted Bi-2201 bilayer on a substrate of monolayer Bi$_2$O$_2$Se. The highlighted bands are degenerate in the absence of SOC, indicating that they are a Rashba pair. These bands are primarily localized to the substrate but have a large component in the bilayer, as shown in Fig.~\figref{structure}{b}.}
        \label{bilayerBands}
\end{figure}

The band structure of this heterostructure revealed the presence of a pair of bands, highlighted in Fig.~\ref{bilayerBands}, which display a large Rashba spin-orbit effect. This can be confirmed by computing the in-plane spin texture of its constituent states. In addition, as Fig.~\figref{structure}{b} illustrates, the wavefunctions of these states are mostly localized to the substrate but contain a substantial tail that extends into the bilayer as well. This indicates a large coupling energy $\Gamma$ between these substrate bands and the bilayer, which in turn enhances the expected proximity-induced pairing energy.

Since our proposal requires the Fermi energy to be moved as close as possible to the bottom of the substrate bands (or, accounting for SOC, to the crossing point at $\Gamma$), we only need to consider the band structure very close to the crossing. By fitting these bands at small momentum to $\epsilon_{\bm{k}}$ and $\mathcal{H}_\text{SOC}$ from Eqs.~\eqref{Hamiltonian_bilayer} and~\eqref{ek}, we obtain $\mu = 497$~meV, $t = 132$~meV and $\alpha = 76$~meV. The last value in particular is highly promising; this relatively large spin-orbit energy, which can be attributed to the high mass of the system's Bismuth atoms, contributes approximately linearly to the induced superconducting gap. As we show in the next section, the parameters associated with these bands are likely to result in a TSC with a measurable gap.

 The only challenge in using these bands is that moving the Fermi energy to the bottom of the band requires gating the system by $\mu = 497$~meV. As a 2D semiconductor with a low carrier density \cite{Wu2017}, the Bi$_2$O$_2$Se substrate monolayer is highly susceptible to having its Fermi energy shifted via an electric field between a top and bottom gate. If, however, this large of a shift proves prohibitive, another combination of cuprate bilayer and substrate could be considered. The value of $\alpha$ demonstrated by the Bi-2201-Bi$_2$O$_2$Se interface suggests that similar Bi-containing cuprate heterostructures can also give rise to the large Rashba energies necessary for a measurable induced TSC gap.

\section{Predicted Induced TSC Gap}
\label{SC}

Now that we have determined the values of all relevant band structure parameters, we can compute the expected gap induced by the superconducting bilayer into the substrate. The only values in our model that must still be determined are the external magnetic field $H_z$ - this will depend on experimental constraints, but as we will see, fields of magnitude larger than 2T will not be needed - and $\Delta_{\bf k}$, the momentum-dependent pairing induced in the substrate.

A substrate in proximity to a $d$-wave SC inherits a pair potential with the same $d$-wave symmetry, as has been observed experimentally in a graphene-cuprate heterostructure \cite{Perconte2020}. This can also be shown analytically through a straightforward generalization of the derivation of the $s$-wave SC proximity effect (see Appendix \ref{derivation} for details). Specifically, the induced pairing $\Delta_k$ is related to the pairing in the superconductor, $\Delta_k^{\text{SC}}$, by
\begin{equation}
\label{deltaind}
    \Delta_{\bm{k}} = \frac{g^2}{e_{\bm{k}}^2 + (\Delta^{\text{SC}}_{\bm{k}})^2} \Delta^{\text{SC}}_{\bm{k}},
\end{equation}
where $e_{\bm{k}}$ is the energy difference between the substrate band and superconducting band (before pairing) and $g$ is the coupling between the SC and substrate. This equation is valid in the limit of small coupling, $g \ll e_{\bm{k}}$.

Since the Fermi pocket is small in our model, and $e_{\bm{k}}$ and $g$ are both finite at the $\Gamma$ point, we can approximate the coefficient of $\Delta^{\text{SC}}_{\bm{k}}$ as a constant:
\begin{equation}
\label{kappa}
    \Delta_{\bm{k}} = \kappa \Delta^{\text{SC}}_{\bm{k}},
\end{equation}
Here, $\kappa$, the proximity ratio, is a number between 0 and~1, where 0 represents no coupling and 1 represents perfect coupling. We will not attempt to derive the value of~$\kappa$, since our model cannot tell us the value of~$g$, and regardless, a one-band derivation following Eq.~\eqref{deltaind} would not necessarily encompass all possible interactions in our many-band heterostructure. The crucial point, however, is that the induced order parameter retains the bilayer's $d+id$-wave structure. We expect $\kappa$ to be relatively large since the substrate states extend into the bilayer (Fig.~\figref{structure}{b}), which suggests strong coupling. However, $\kappa$ will also depend on factors such as disorder at the boundary that are not present in our model.

\begin{figure}
     \includegraphics[width=0.45\textwidth]{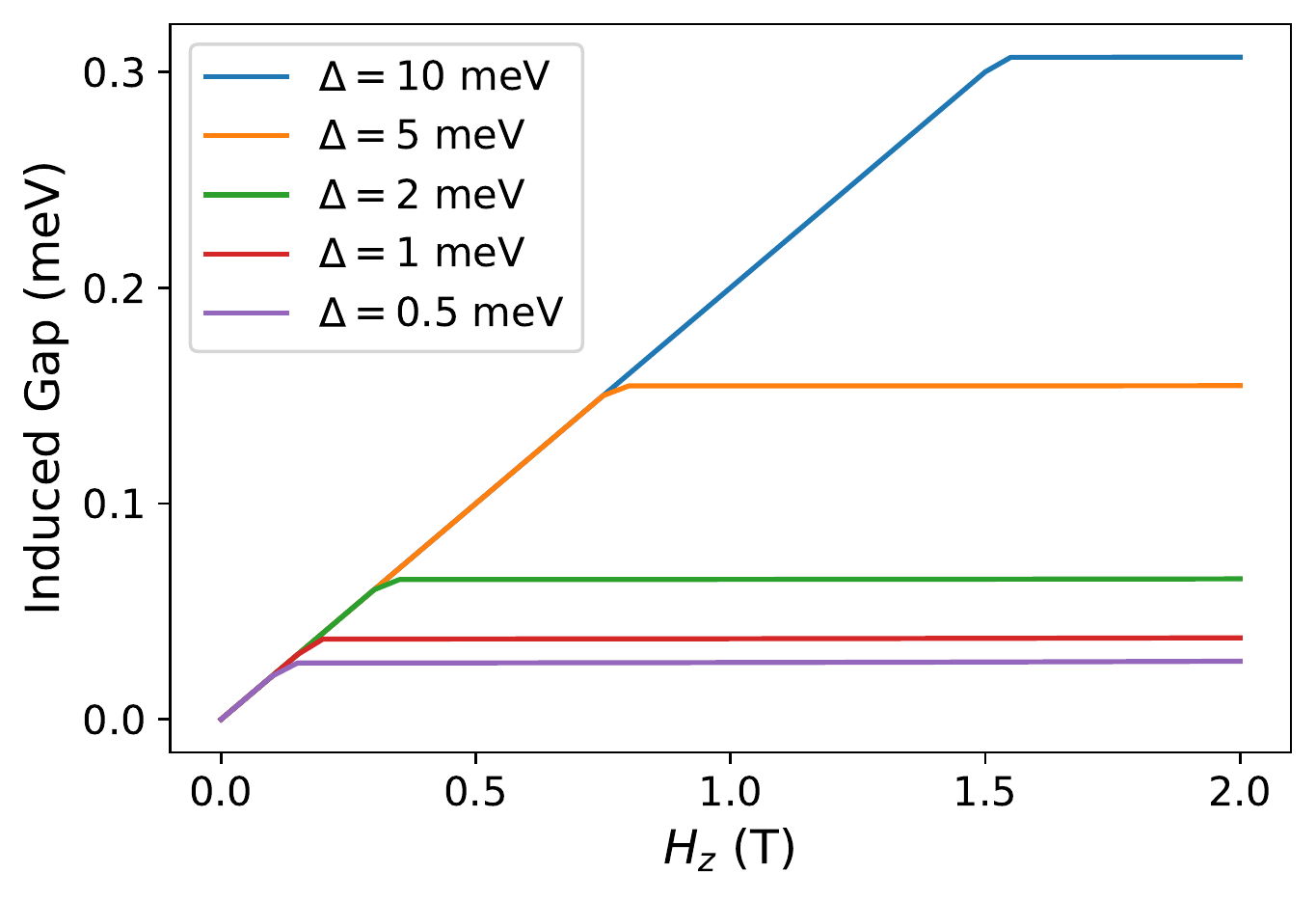}
        \caption{Induced TSC gap as a function of the applied magnetic field $H_z$, for various possible values of the induced maximum pair potential $\Delta$. Increasing $H_z$ yields a larger TSC gap, but this saturates at large field to a value proportional to $\Delta$.}
        \label{TSCgap}
\end{figure}

In the best-case scenario, the induced pairing $\Delta$ is 12~meV if $\kappa = 1$ and the maximal intrinsic gap in the bilayer, $\Delta^{\text{SC}}$, is~12~meV, the same value as bulk Bi-2201 (in the similar material Bi-2212, monolayer pairing has been shown to be nearly identical to bulk pairing \cite{Yu2019}). With this as the upper bound, we can plot the full induced amplitude of the pairing gap (Fig.~\ref{TSCgap}) as a function of applied magnetic field $H_z$ at multiple possible values of $\Delta$.

The energy gap $\delta E$ increases linearly with $H_z$, then levels off. This is because the BdG gap has two possible minima: one centered at $\Gamma$ that depends on $H_z$ as $\delta E=2\mu_B H_z$ (the gap is smaller if the gating is not tuned such that the crossing at $\Gamma$, highlighted in Fig.~\ref{bilayerBands}, lies exactly at the Fermi energy), and another proportional to $\Delta$. The size of the gap is the smaller of these two values, and therefore may be limited either by the applied field or the coupling between the substrate and bilayer.

In the plateaus, the equation for the gap has 3 different reduction factors, as follows:

\begin{equation}
\label{gap1}
    \delta E = \frac{1}{2}\kappa \bigg( \frac{\alpha}{t} \bigg)^2 \bigg( \frac{a_0}{a} \bigg)^2 \Delta^{\text{SC}}.
\end{equation}

This equation is a small-momentum approximation ($ka \ll 1$) and also uses the fact that $\mu_B H_z$ and $\Delta$ (the proximity-induced pairing in the substrate) are both much smaller than $t$ (the substrate bandwidth) and $\alpha$ (the substrate Rashba energy).

The first factor, $\kappa$, is due to the proximity effect, as was discussed earlier in this section. The second, $\big( \frac{\alpha}{t} \big)^2$, is dependent on the material properties of the substrate, and illustrates why a large Rashba energy is crucial for this candidate system. Finally, $\big( \frac{a_0}{a} \big)^2$ is due to the fact that band folding in the moir\'e supercell reduces the scale of $k$ by a factor of $\frac{a_0}{a}$, where $a_0$ is the lattice constant of a Bi-2201 monolayer and $a$ is the moir\'e lattice constant. At small momentum, $d$-wave order parameters are proportional $k^2$, hence the square lattice constant ratio. In our simulation, there are 5 unit cells per moir\'e supercell, so $\big( \frac{a_0}{a} \big)^2 = \frac{1}{5}$.

This last factor does not imply that $\delta E$ decreases as the moir\'e cell increases in size, however. Eq.~\eqref{gap1} implicitly uses $\alpha$ and $t$ (spin orbit energy and bandwidth, respectively) corresponding to the moir\'e cell, but in fact, due to how these parameters scale with lattice constant, the moir\'e scaling exactly cancels out in Eq.~\eqref{gap1}. That is, $\big( \frac{\alpha}{t} \big)^2 \big( \frac{a_0}{a} \big)^2$ is independent of $a$ at this order. Thus, we expect our conclusions to hold even if the twist angle deviates slightly from $36.9^\circ$.

Using 1 meV as a reasonable estimate for $\kappa\Delta^{\text{SC}}$ (\textit{i.e.} the cleaving of the bilayer and imperfect coupling together contribute an order-of-magnitude reduction in pairing from the intrinsic value of 12 meV for bulk Bi-2201), we then apply the following reduction factors from Eq.~\eqref{gap1}: the coefficient of $\frac{1}{2}$, roughly $\frac{1}{3}$ from the $\big( \frac{\alpha}{t} \big)^2$ factor, and $\frac{1}{5}$ from the moir\'e factor $\big( \frac{a_0}{a} \big)^2$, for a total of about $\frac{1}{30}$. We arrive at a net TSC gap of $\delta E=0.033$~meV. This energy gap is large enough that it may be measurable in experiment. At sufficiently-low temperatures, we thus expect the system to behave as a Chern-1 (or Chern-$(-1)$) TSC, and hence to observe chiral Majorana modes on the system boundary and Majorana zero modes in vortices.

\section{Conclusion}
\label{Conclusion}

We have shown that the $d + id$-wave superconductivity of a twisted bilayer formed from a high-$T_C$ SC material can yield a chiral TSC when placed on a substrate in a magnetic field. This relies on a pair of bands with Rashba SOC and a Fermi level close to their intersection at the $\Gamma$ point.

Due to the compounding effect of reduction factors as illustrated in Eq.~\eqref{gap1}, it is challenging in general for chiral TSC heterostructure candidates to yield large pairing energies. Our proposal meets this challenge by leveraging the large intrinsic pairing energies of the high-$T_C$ superconductors, as well as a unique $d+id$-wave order parameter which can yield a chiral TSC for arbitrarily low Zeeman energies. 

The predicted induced SC gap is within observable ranges for plausible coupling energies and magnetic field strengths, making this class of heterostructure a contender for realizing the chiral TSC phase experimentally.

\section*{Acknowledgements}
G.M. wishes to thank Ryan Day for helpful discussions concerning DFT simulations of the bilayer. B.Y. acknowledges financial support by the European Research Council (ERC Consolidator Grant ``NonlinearTopo'', No. 815869) and the MINERVA Stiftung with the funds from the BMBF of the Federal Republic of Germany. M.F. research was supported in part by NSERC and the Canada First Research Excellence Fund, Quantum Materials and Future Technologies Program. Y.O acknowledges the European Union’s Horizon 2020 research and innovation programme (Grant Agreement LEGOTOP No. 788715), the DFG (CRC/Transregio 183, EI 519/7-1), ISF Quantum Science and Technology (2074/19), the BSF and NSF (2018643).

\appendix
\section{Derivation of $d + id$-wave Proximity Effect}
\label{derivation}

Here we prove our assertion that a $d + id$-wave SC can induce $d + id$-wave SC gap in a substrate via the proximity effect, and that the specific form of this gap is given by Eq.~\eqref{deltaind}. We begin with a simple model of a substrate $\mathcal{H}_0$ with fields $\psi_{\bm{k}s}$ - where s is spin and $\bm{k}$ is momentum - coupled to a superconductor $H_{\text{SC}}$ with fields $\eta_{\bm{k}s}$:

\begin{equation}
\label{Hamiltonian_derivation}
\begin{aligned}
\mathcal{H} &= \mathcal{H}_{\text{0}} + \mathcal{H}_{\text{SC}} + \mathcal{H}_{g}; \\
\mathcal{H}_{\text{0}} &= \sum_{\bm{k},s}{ \psi_{\bm{k}s}^\dagger (\mathrm{H}_{\bm{k}})_{s,s'} \psi_{\bm{k}s'}} \\
\mathcal{H}_{\text{SC}} &= \sum_{\bm{k},s}{e_{\bm{k}}\eta_{\bm{k}s}^\dagger \eta_{\bm{k}s}} \\
&+ \sum_{\bm{k}}{ \Delta_{\bm{k}}^{\text{SC}}\eta_{\bm{k}\uparrow}^\dagger \eta_{-\bm{k}\downarrow}^\dagger - (\Delta_{\bm{k}}^{\text{SC}})^* \eta_{\bm{k}\uparrow}\eta_{-\bm{k}\downarrow}} \\
\mathcal{H}_{g} &= -g\sum_{\bm{k},s}{ \psi_{\bm{k}s}^\dagger \eta_{\bm{k}s} + \text{H.c.}},
\end{aligned}
\end{equation}
where $\mathrm{H}_{\bm{k}}$ is the substrate Hamiltonian matrix, which may include SOC and magnetic field terms, $e_{\bm{k}}$ is the band structure of the superconductor, $\Delta_{\bm{k}}^{\text{SC}}$ is its pair potential, and $g$ is the coupling energy between the superconductor and substrate. We assume that the coupling is momentum-independent and spin-independent, and hence that $\eta_{\textbf{k}s}$ couples only to $\psi_{\textbf{k}s}$ with the same $\bm{k}$ and $s$.

First, we write $\mathcal{H}_{\text{SC}}$ as
\begin{eqnarray}
\label{BdG}
\mathcal{H}_{\text{SC}} = \frac{E_{\bm{k}}}{2}
\begin{pmatrix}
x_{\bm{k}} & y_{\bm{k}}^*  \\
y_{\bm{k}} & -x_{\bm{-k}}
\end{pmatrix},
\end{eqnarray}
where $E_{\bm{k}} = \sqrt{e_{\bm{k}}^2 + \abs{\Delta_{\bm{k}}^{\text{SC}}}^2}$, $x_{\bm{k}} = \frac{e_{\bm{k}}}{E_{\bm{k}}}$, and $y_{\bm{k}} = \frac{\Delta_{\bm{k}}^{\text{SC}}}{E_{\bm{k}}}$. We then diagonalize $H_{\text{SC}}$:
\begin{equation}
\label{chi_def}
\begin{aligned}
\chi_{1\bm{k}} &\equiv u_{\bm{k}} \eta_{\bm{k}\uparrow} + v_{\bm{k}} \eta^{\dagger}_{\bm{-k}\downarrow}\\
\chi^{\dagger}_{2\bm{-k}} &\equiv v^*_{\bm{k}} \eta_{\bm{k}\uparrow} - u_{\bm{k}} \eta^{\dagger}_{\bm{-k}\downarrow},
\end{aligned}
\end{equation}
where $u_{\bm{k}} = \sqrt{\frac{1}{2}(1 + x_{\bm{k}})}$ and $v_{\bm{k}} = \frac{y_{\bm{k}}}{\abs{y_{\bm{k}}}}\sqrt{\frac{1}{2}(1 - x_{\bm{k}})}$.

In this new basis,
\begin{equation}
\label{HSC_diag}
\mathcal{H}_{\text{SC}} = \sum_{\bm{k},s}{E_{\bm{k}}(\chi^{\dagger}_{1\bm{k}}\chi_{1\bm{k}} + \chi^{\dagger}_{2\bm{k}}\chi_{2\bm{k}})}.
\end{equation}

We solve for $\eta_{\bm{k}s}$ in terms of $\chi_{1\bm{k}}$ and $\chi_{2\bm{k}}$ and substitute into $\mathcal{H}_g$ to obtain

\begin{equation}
\label{Hg}
\begin{aligned}
\mathcal{H}_{g} &= -g\sum_{\bm{k}}{ \psi_{\bm{k}\uparrow}^\dagger (u_{\bm{k}} \chi_{1\bm{k}} + v_{\bm{k}} \chi^{\dagger}_{2\bm{-k}})} \\
&\quad\quad\quad\; + \psi_{\bm{k}\downarrow}^\dagger (v_{\bm{-k}} \chi^{\dagger}_{1\bm{-k}} - u_{\bm{k}} \chi_{2\bm{k}}) + \text{H.c.}\\
&= -g\sum_{\bm{k}}{(u_{\bm{k}}\psi_{\bm{k}\uparrow}^\dagger - v_{\bm{k}}\psi_{\bm{-k}\downarrow})\chi_{1\bm{k}}} \\
&\quad\quad\quad\; + (-v_{\bm{-k}}\psi_{\bm{-k}\uparrow} - u_{\bm{k}}\psi_{\bm{k}\downarrow}^\dagger)\chi_{2\bm{k}} + \text{H.c.}
\end{aligned}
\end{equation}

We now complete the square, making a Hubbard-Stratonovich transformation to find the effective pairing caused by this proximity-induced interaction with the superconductor. This relies on the assumption that $g/E_{\bm{k}}$ is small; the resulting effective pairing is second-order in this ratio.

Ignoring the diagonal terms, which modify the particle or hole band structure but do not contribute to pairing, the additional square term introduced by the transformation is
\begin{equation}
\label{deltaHg}
\delta \mathcal{H}_g = -\frac{g^2}{E_{\bm{k}}}\sum_{\bm{k}}{ 2 u_{\bm{k}} v_{\bm{k}} \psi_{\bm{k}\uparrow} \psi_{\bm{-k}\downarrow} + \text{H.c.}}.
\end{equation}
Thus, the induced pairing is
\begin{equation}
\label{deltaind2}
\begin{aligned}
\Delta_{\bm{k}} &= \frac{2g^2}{E_{\bm{k}}} u_{\bm{k}} v_{\bm{k}} \\
&= \frac{g^2}{E^2_{\bm{k}}} \Delta^{\text{SC}}_{\bm{k}},
\end{aligned}
\end{equation}
from which we arrive at Eq.~\eqref{deltaind}.

This denomenator is dominated by $e_{\bm{k}}$, which, for small $\bm{k}$, we can assume to have a quadratic dispersion plus a constant offset from the Fermi energy. Since this function is isotropic, it does not change the symmetry; thus, if $\Delta^{\text{SC}}_{\bm{k}}$ is $d+id$-wave, then the proximity-induced pairing $\Delta_{\bm{k}}$ will have a $d+id$-wave character as well. 

\bibliographystyle{aipnum4-2}
%

\end{document}